\documentclass[twoside]{article}
\usepackage[T1]{fontenc}
\usepackage[a4paper, total={6.5in, 9in}]{geometry}

\usepackage{graphicx,sepnum,xspace,bbding}
\usepackage{hyperref}
\usepackage{fancyhdr}
\usepackage{comment}
\usepackage{xcolor}
\usepackage{color}
\usepackage{fvextra}
\usepackage{wrapfig}
\usepackage{tabularx}

\newcommand{\point}[1]{\par\smallskip\noindent\textbf{#1}.}
\newcommand{\code}[1]{\Verb{#1}}

\usepackage{siunitx}

\usepackage{amsmath}
\usepackage{amsfonts}
\usepackage{nicefrac}

\usepackage[labelfont = bf,font=footnotesize]{caption}
\usepackage[labelformat=simple]{subcaption}

\newcommand{\etal}{\textit{et al.}}

\usepackage[numbers,sort&compress, sectionbib]{natbib}

\usepackage{acronyms}
\usepackage{hyperref}
\hypersetup{
    colorlinks=true,
    linkcolor=blue,
    filecolor=magenta,      
    urlcolor=cyan,
    pdftitle={Overleaf Example},
    pdfpagemode=FullScreen,
    }

\begin{document}
\sloppy
%
\title{Maxing Out the SVM: Performance Impact of Memory and Program Cache Sizes in the Agave Validator}
%
%
\author{Turan Vural \and Yuki Yuminaga  \and Alex Petrosyan \and Benjamin Livshits}
%
\maketitle              

\begin{abstract}
In this paper we analyze some of the bottlenecks in the execution pipeline of Solana's Agave validator client, focusing on RAM and program cache usage under mainnet conditions. Through a series of controlled experiments, we measure the validator’s throughput and resource efficiency as RAM availability ranges between~128~GB to~1,536~GB~(1.5 TB). We discover that the validator performance degrades significantly below~256~GB, with transaction processing falling behind real-time block production. Additionally, we study the program cache behavior, identifying inefficiencies in program eviction and load latency. Our results provide practical guidance for hardware provisioning and suggest improvements to the Solana execution and caching strategy, reducing latency due to the program cache by~90\%.

\end{abstract}

\section{Introduction}%
\label{sec:intro}

The Solana blockchain is famous for its high hardware requirements for validator nodes. 
In this article, we take a look at two potential bottlenecks of execution in Agave, an implementation of the Solana validator: RAM and the program cache. 
Specifically we quantify RAM required to for a validator to remain in lock-step with the mainnet-beta\footnotemark.
We do this by comparing transaction throughput, measured in Transactions Per Second (TPS) as well as other memory-specific metrics, that together paint a complete picture of the impact on the validator's performance.
We also investigate the program cache used by the Solana program JIT, characterizing its behavior and latency in the current validator, while running experiments to make the cache more efficient in execution.
\footnotetext{%
Solana uses the same transactions and programs deployed in a low-risk environment, wherein some rollback is possible, called the mainnet beta. This environment is otherwise identical to the main network in both transaction load and types of transactions.  As such, the results from mainnet-beta are transferable to the mainnet.%
}

\subsection{Paper Organization}
The rest of the paper is organized as follows.
In Section~\ref{sec:background} we give a brief overview of Solana-specific concepts and the processing pipeline. 
In Section~\ref{sec:execution} we talk specifically about the execution pipeline, and the SVM. 
In Section~\ref{sec:programcache} we discuss the program cache, its role in the SVM and the effects of allocating insufficient RAM to it. 
Section~\ref{sec:related-work} provides an overall comparison of our work to previous work. 
The conclusions are given in Section~\ref{sec:conclusion}. 

\section{Background}
\label{sec:background}

In this section, we give an overview of Solana's execution model and the program cache.
We first focus on parallel execution, then we explain the four stages of execution in the Solana validator: the TPU, the Bank, the SVM, and the checks that need to be performed on a program before it reaches the eBPF VM. We then explain the role and functionality of the program cache, along with some of the optimizations necessary for a blockchain-oriented program cache.

The Transaction Processing Unit~(TPU), is the component of the validator responsible for block production. 
The TPU ingests transactions from the network, executes them in batches, and assembles them into blocks for verification.
The TPU consists of four stages:
\begin{enumerate}
\item{Fetch Stage}
\item{\code{SigVerify} Stage}
\item{Banking Stage}
\item{Broadcast Stage}
\end{enumerate}

The Fetch stage is responsible for ingesting transactions when the current node is the \emph{leader} in the current consensus topology.
The \code{SigVerify} stage is responsible for validating the transactions.
Specifically, it confirms that the (usually ED25519) signature matches the message and public key of the signer and removes the duplicate and expired transactions.

At this point, the transaction is ready to be executed in the banking stage.
It is named after the Bank, a data structure that maintains Solana's account state, tracks account balances and executed transactions, deducts rent and consumes a necessary amount of compute units, which is Solana's equivalent of gas in EVM-based systems. SVM programs are stateless, in that they can only read from a set of accounts that is declared ahead of time.
As such, the Solana programs interact with the Bank via an API.
This enforces multiple blockchain state invariants, \emph{e.g.} that the transactions had been executed against the correct account state. The banking stage is also responsible for the final step in the validation of transactions. Transactions modify the Bank state, which is then finalized into a block and broadcast to the network.

The SVM is a parallel execution environment.
Unlike Ethereum, which forces all transactions to execute one at a time, SVM groups transactions into batches that can be executed concurrently.
To this end, each transaction contains an exhaustive list of both input data (readable accounts) and possible state modifications (writable accounts).
The Bank's scheduler analyzes these dependencies to determine which transactions can be executed without conflicts.
Specifically, only read dependencies are allowed to overlap other read dependencies, while write dependencies are not allowed to overlap with either read or write dependencies.
From these transactions a batch is formed, which is executed using multiple threads.
This setup allows a high degree of parallelism: in principle, transactions in every batch can be executed concurrently.
By default Agave uses four execution threads. 
This number is chosen for historical reasons, as well as to offer a good middle-ground between fast execution core utilization.
In practice validators often raise the number of allocated threads.
The effects of the execution are committed back to the bank and forwarded to the next stage.

At a point where a block is sufficiently populated with transactions, and/or sufficient time has elapsed, a bank is ``frozen'', meaning that it does not allow any further modifications.
This frozen bank is what then becomes the next block.
It is committed to persistent storage and broadcast to a component called \code{turbine}.

Having discussed the ingest/execute/publish cycle, it is worth discussing some of the validity checks applied to the transactions.
\begin{enumerate}
\item \textbf{Blockhash expiry check}.
Every Solana transaction references a recent block hash. If the block hash is too stale, the transaction is considered expired and is not executed. This puts a deterministic upper bound on the lifetime of a transaction.

\item \textbf{Precompile verification}.
Solana offers the ability to execute pre-compiled programs, meaning native code that exists outside of the eBPF runtime.
The pre-compiles available can vary with validator versions, as new ones can be introduced, and older ones deprecated and retired.
To ensure no memory access violations as well as protect against undefined behaviour, the system verifies that any referenced precompiled program is valid and enabled%
\footnote{%
Solana allows decoupling on-chain state changes from binary upgrades of the validators.
Thus, for example, a new precompiled program can be introduced in version~1.16, enabled in 1.17 and made mandatory in version~1.18.}.

\item \textbf{Transaction costs and resource constraints}.
Most blockchains meter the execution of smart contracts.
This is called \emph{gas} in the Ethereum parlance,  and \emph{Compute Units} in Solana.
Every transaction's execution cost must be within a set of global limits.
The scheduler tries to balance the execution between blocks, such that the total number of compute units is close to, but less than the per-block CU limit.
Additionally, each program has a separate limit that it needs to respect, to prevent spam and denial of service.

\item \textbf{Account locking and block limits}.
The Bank then attempts to read-write lock all accounts involved in the transaction.
If the required accounts are already locked by another transaction, the transaction may be deferred or re-tried in a later block.
The system also ensures that the transaction does not exceed any of the block-level constraints, including:
\begin{itemize}
    \item \code{AccountInUse}: an account needed by the transaction has been locked by another thread.

    \item \code{WouldExceedAccountDataBlockLimit}: the transaction exceeds the maximum data limit for the block.

    \item \code{WouldExceedMaxBlockCostLimit}: the transaction exceeds the maximum compute cost of the block.

    \item \code{WouldExceedMaxAccountCostLimit}: the transaction exceeds the maximum number of accounts that can be locked by the block.

    \item \code{WouldExceedMaxVoteCostLimit}: the transaction exceeds the maximum vote cost for the block.

    \item \code{TooManyAccountLocks}: the transaction tries to lock too many accounts.
\end{itemize}

\end{enumerate}
Transactions that pass all of the above checks are called \emph{sanitized transactions}, and are considered safe to execute.
The execution in the eBPF runtime can still report an error, \emph{e.g.} because some of the program's invariants have been violated, or because the transaction does not respect blockchain invariants.
Failure to satisfy any of the above constraints (except \code{TooManyAccountLocks}), results in the transaction being re-tried in a later batch, until it either succeeds or expires.

\subsection{Program Cache}
The transactions are executed by loading a specific account's data, known as the program owner, which is executable eBPF\footnotemark code. To avoid unnecessary recompilation, a cache is used; this cache is specific to Solana, as it must account for blockchain forks, ensuring that execution state remains consistent across multiple potential branches of the ledger.
\footnotetext{Technically, the version of eBPF used in Solana, called SBPF, is different from the ``main-line'' eBPF that is embedded in \emph{e.g.} the Linux kernel}
 
The Solana validator leverages a multi-layered hierarchy of abstractions to manage data access during transaction execution.
We illustrate the flow of a typical transaction execution in the SVM:
\begin{enumerate}
\item \textbf{Cache hit}: The SVM first checks the Program Cache for the compiled program's bytecode and the Accounts Cache for account data.
  Both caches reside in RAM, and bytecode is interpreted or compiled by JIT into the validator's native machine code for optimal performance.
\item \textbf{Cache miss}: If the required data are not in the cache, they are fetched from the corresponding databases: the Ledger for program instructions, Accounts for account state.
  All this information is stored on high-performance NVMe disks.
  Once data is fetched, they are placed in the cache.
\item \textbf{Execution}: The cached data are passed to the SVM, which executes the transaction logic.
\end{enumerate}

Programs are pre-loaded before batch execution begins, enabling collaborative execution for transactions that invoke the same program within a batch.
The cache is only populated at batch boundaries --- lazy loading during batch execution is not supported, nor is any form of ``look-ahead'' optimization currently implemented.
Thus, the program cache is not tied to any specific snapshot or chain state and must be warmed up on startup.

A further constraint is that it must track separate program state across blockchain forks and between batches.
Namely, forks introduce different states for the ledger, and the program cache must account for these state variations across branches.
This requirement imposes a hierarchy upon objects in the program cache that distributes loaded programs across multiple layers of abstraction.

The cache handles re-rooting: when a branch is orphaned, the validator prunes the cache to free up memory.
Additionally, the cache recognizes Tombstones, which represent programs that are invalidated due to verification failures or other specific reasons.
Tombstones prevent unnecessary reloading of invalid programs.

The program cache is global.
This allows it to serve any valid state corresponding to any fork managed by the validator.
This design facilitates sharing, which reduces data duplication.
For instance, programs with identical states on different forks can share a single cache entry.

Unlike traditional caches, the program cache is \emph{elastic}, with a target size of~256 (increased to~512 in~v2.1) loaded programs when warm.
The program cache tracks load and usage statistics at the \code{ProgramCache} level, using these to implement a Least Frequently Used~(LFU) eviction strategy.
However, instead of evicting the~10\% of the LFU programs, two programs are chosen at random, and the one with a lower frequency is evicted, until the size of the cache is reduced to~90\% of the capacity.
This is done to prevent attacks targeting the program cache from affecting more than one validator.
Additionally, random 2 eviction offers little overhead in exchange for acceptable eviction efficiency.
An alternative such as a full-blown LFU would be much harder to maintain and offer a bigger attack surface due to the low diversity of most frequently used programs.


\section{Execution}
\label{sec:execution}

Solana validators process transactions at high throughput, requiring significant memory resources to cache program execution, manage accounts, and reduce disk access latency.
Understanding how RAM availability impacts validator performance is crucial for optimizing validator deployments.
To evaluate the impact of memory availability on Solana validator's performance, we conducted a series of experiments running the Agave validator with varying amounts of RAM, ranging from~128~GB up to~1,536 GB~(1.5~TB).
\begin{wrapfigure}{r}{0.45\textwidth}
    \centering\footnotesize
    \setlength{\tabcolsep}{2pt}
    \begin{tabular}{lr}
    \toprule
    AWS configuration & \code{r6a.metal}\\
    \midrule
    CPU	&   AMD EPYC 7R13 \\
    Core count & 192  \\
    Base CPU clock & 3.6 GHz, x86\_64 \\
    \midrule
    Memory	& 1,536~GiB \\
    \midrule
    Network	& 40~GiB/s \\
    \bottomrule
    \end{tabular}
    \caption{The CPU configuration of the test machine}
    \label{fig:cpu-hardware}
\end{wrapfigure}
Our goal was to determine the minimum viable memory required to keep up with the chain; to find performance breakpoints which provide optimal transaction throughput~(TPS) and to evaluate the impact of operating a node with sub-optimal memory allocation.

In this section, our aim is to answer the following questions.
\begin{enumerate}
    \item What is the minimum amount of RAM needed for execution?
    \item How does memory and execution behave as more memory is made available to the validator process?
\end{enumerate}

\subsection{Methodology}
Figures~\ref{fig:cpu-hardware} and \ref{fig:nvme-hardware} show the test hardware that was used for our experiments. 
We used \code{agave-validator v2.0.5} to run our non-voting validator on mainnet-beta, devnet, and testnet.
At the time these experiments were conducted, this was the latest version of Agave available.
Presently, many of the issues that we have identified have been both reported and fixed.

A systemd service was used to automate experiments and control system resources.
Specifically: to set the user and appropriate environment variables as well as
call the agave validator startup script.

\begin{wrapfigure}{r}{0.4\textwidth}
    \centering\footnotesize
    \setlength{\tabcolsep}{2pt}
    \begin{tabular}{l r}
    \toprule
    Root drive (NVMe)\\
    \midrule
    Size & 512 GiB\\
    EBS type & gp3 \\
    IOPS & 3,000 \\
    Throughput	& 250 MiB/s\\
    \midrule
    Ledger drive (NVMe)\\
    \midrule
    Size		& 2048 GiB\\
    EBS type	& gp3\\
    IOPS		& 10,000\\
    Throughput	& 700 MiB/s\\
    \bottomrule
    \end{tabular}
    \caption{Drive layout and performance characteristics for the test machine.}
    \label{fig:nvme-hardware}
\end{wrapfigure}

As such, we specified bootstrapping behavior, trusted endpoints, and included commands to clear the ledger.
We then started the memory management script.
We waited for the \code{agave-validator} process to start, then tracked system resource usage by process ID.

Each experiment was run for a maximum of~12 hours, or until an out-of-memory failure occurred.
RAM limitations were tested by running the validator with~256~GB, 512~GB, 1,024~GB, and~1,536~GB of RAM.

The experiments were orchestrated with a \code{systemd} file. We ran experiments as a validator that did not participate in consensus (unstaked) on mainnet-beta, devnet, and testnet with varying hardware configurations. All experiments were run for a maximum time of~12~hours, or until an out-of-memory failure occurred. The ledger was cleared at the start of every experiment, requiring the node to download a snapshot and catch up to the tip of the chain.

Our intention is to capture real-world performance data under main network conditions, including:
\begin{itemize}
    \item \textbf{Leader shred ingestion}: validators process data streamed from the network.
    \item \textbf{Variable transaction throughput}: TPS fluctuates dynamically based on network congestion.
    \item \textbf{Snapshot-based bootstrapping}: validators must catch up to the tip of the blockchain after every restart.
    \item \textbf{Ledger pressure}: real-world execution includes database writes, memory allocations, and eviction policies under realistic workloads.
\end{itemize}

\subsection{Results}
In this section, we present the effects of available RAM on the physically utilized memory: RSS,  total virtual memory: VSZ, , page faults, minor and major; as well as their cumulative effects on the node's transaction throughput.
We have identified certain efficiency breakpoints.  
For instance,~512~GiB is an optimal memory configuration, allowing stable throughput and sufficient headroom for chain stability. 
The~128~GiB memory configuration demonstrated inability to keep up with the main network.
The most common memory configuration:~256~GiB we identified as having acceptable performance.

We begin our discussion with Figure~\ref{tab:ram-performance}, which shows the key statistics regarding transaction throughput.

\begin{figure}[tb]
\centering\footnotesize
\setlength{\tabcolsep}{5pt}
\begin{tabular}{r c r r r}
\toprule
\textbf{RAM} & \textbf{Date} & \textbf{Avg TPS} &  \textbf{0 TPS Time} & \textbf{Max TPS} \\
\midrule
1.5~TB & 01/07/2025 & \num{4598.12}  & \num{10039.3} & \num{120790.77} \\

1.5~TB & 01/01/2025 & \num{4742.27}  & \num{8428.4} & \num{68038.50} \\

1~TB   & 01/07/2025 & \num{5078.81}  & \num{9992.8} & \num{126476.59} \\

1~TB   & 01/04/2025 & \num{4634.92}  & \num{10640.1} & \num{89009.14} \\

512~GB & 01/08/2025 & \num{4642.67} & \num{13492.9} & \num{122269.14} \\

256~GB & 01/09/2025 & \num{4486.91} &  \num{15513.4} & \num{116547.52} \\

128~GB\footnotemark & 01/09/2025 & \num{0.00}    & \num{43200.0} & N/A \\
\bottomrule
\end{tabular}
\caption{Validator performance across RAM configurations and test runs. The~0~TPS time is the time in which the validator had reported zero transactions processed.}
\label{tab:ram-performance}
\end{figure}
\footnotetext{The 128 GiB test did not catch up after snapshot-based bootstrapping.}

\subsubsection{TPS Performance and Catch-up Behavior}

TPS graphs are scaled to capture the majority of execution data, with time measured from the first execution (excluding snapshot download time). Outliers are excluded from the average calculation but included in the maximum TPS metric, where peaks exceed~\(3\times\) the average TPS.
Idle Time as an Indicator of Bottlenecks: To supplement the average TPS, we measured time spent at~0 TPS. Successful validator runs with~1~TB and~1.5~TB~RAM show less~than~3~hours of idle time, while~512~GB and~256~GB runs show significantly longer idle periods.

\point{Chain Conditions vs. Validator Limitations} Given that these experiments were carried out over multiple days (see second column in Fig.~\ref{tab:ram-performance}) with varying network conditions, consistent performance differences between runs indicate that idle execution time and TPS fluctuations are primarily a result of the constraints of the validator resource, rather than changes in chain activity.

\subsubsection{Memory Utilization and Page Fault Behavior}

Tracking Memory Usage: We logged physical memory (RSS: the amount of application memory in RAM) and total memory (VSZ: the amount of memory in RAM and swap) to track how the validator utilized RAM. Additionally, major page faults (requiring disk access) and minor page faults (cache-based) were recorded to understand the memory pressure.

\point{1~TB and~1.5~TB Stability} Both configurations showed a low number major page faults after the initial warm-up phase, and stable memory usage. 
Physical memory use was around~512~GB, with total memory utilization (RSS + VSZ) between~600~GB and~700~GB.

\point{512~GB, first signs of instability} While~512~GB RAM kept up with the chain, a greater reliance on virtual memory led to greater fluctuations in TPS.
This is consistent with the latest recommendations from Anza to disable swap on Linux machines.
The total memory usage (physical + virtual) was approximately~1~TB, significantly exceeding the memory footprint of the~1~TB+ runs. 
Memory saturation and extended warm-up times (1 hour) contributed to increased execution instability.

\point{256~GB Shows Clear Degradation}
This setup failed to keep up with the chain and exhibited a lower TPS floor (3,800 TPS, which is~1,000 TPS short of the needed levels). 
Although it technically completed execution without OOM, its inability to accelerate during bootstrapping and excessive use of virtual memory show clear signs of resource starvation.
These issues have been reported and addressed in the latest versions of Agave.

\point{128~GB Fails with OOM Error} The 128~GB validator did not keep up with the chain and was terminated due to out-of-memory failure less than~3 hours into execution. 
With an average TPS of only~1,200 TPS (compared to~4,500+ for successful runs) and a maximum TPS of just~2,300~TPS, it lacked the necessary resources to sustain execution.

\subsubsection{Discussion}
Our findings indicate that RAM choice has a profound impact on the viability of a full node operator's ability to participate in the Solana mainnet chain.
As of writing the safest choice is to have 512~GiB of physical RAM on the machine running the validator.
The most popular configuration:~256~GiB is shown to be dangerously close to instability.

\begin{figure}
    \centering
    \includegraphics[width=.75\linewidth]{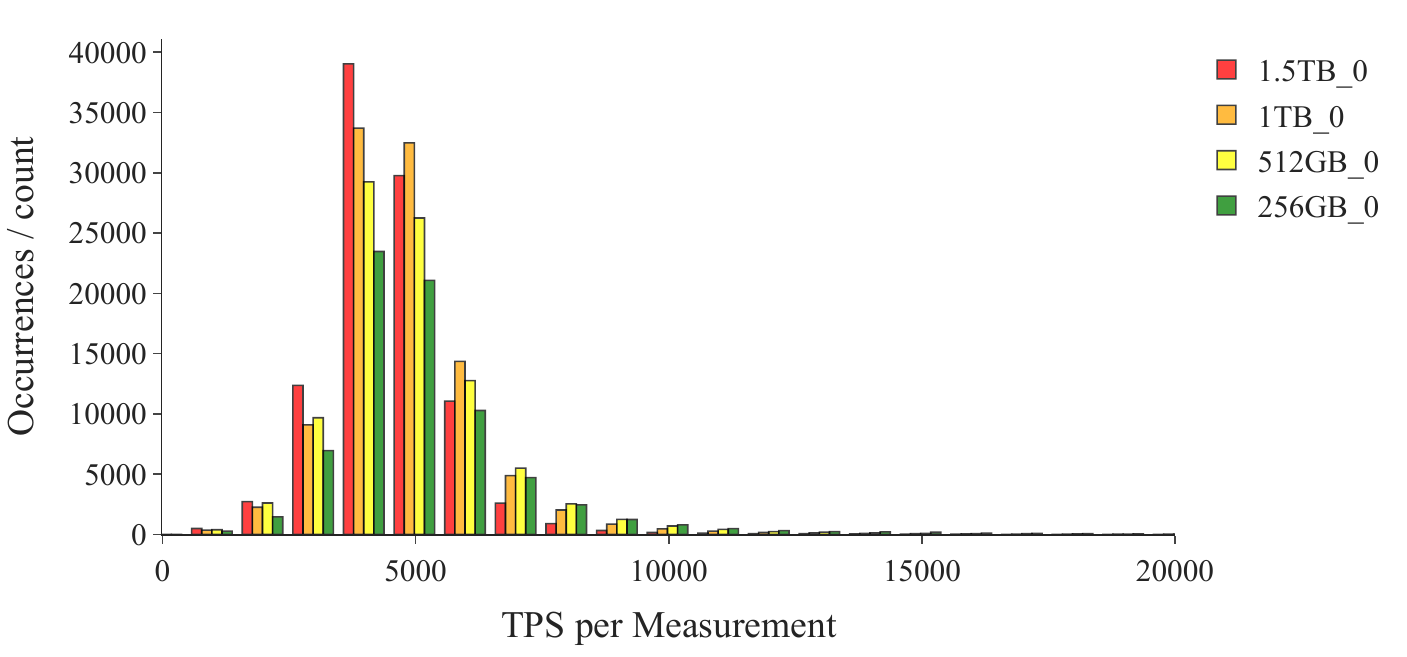}
    \caption{Comparison of TPS instances across validators with varying memory sizes. 
    1.5 TB shows the most instances of TPS corresponding to the average TPS of the validator across the experiment. 
    1 TB shows similar behavior with a skew towards slightly above higher instances. 
    512 GB shows a significant need for catch up, with comparatively more instances of higher TPS and less of average/low TPS, suggesting that it cannot keep up with transaction ingestion. 
    256 GB shows insufficient instances altogether.}
    \label{fig:tps_histogram}
\end{figure}

\point{1.5~TB Demonstrates the Smoothest Execution} The shortest catch-up time (1 hour) and low TPS fluctuation indicate that excess memory reduces execution variance and allows for predictable performance.
The~1~TB  configuration shows similar stability, slightly longer catch-up. 
Specifically, one run had an identical catch-up time (1 hour) to the~1.5TB run, while another took slightly over~2 hours but exhibited extended high-load performance~(12K TPS).

\point{512~GB is the Bare Minimum for Sustained Performance} While it kept up with the chain, TPS stability was significantly worse. It suffered longer idle periods, greater warm-up times, and higher virtual memory reliance, making execution less predictable.

\point{256~GB and Below Cannot Sustain Validator Execution} While~256~GB completed execution, its inability to sustain the required TPS and reliance on virtual memory suggest it cannot function as a performant validator. 

\point{128~GB} This setup was entirely not viable due to an OOM failure.

\section{Program Cache}
\label{sec:programcache}

Following our RAM usage experiments, which indicated that a validator requires approximately~700~GB of RAM to run reliably—and potentially more to fully utilize memory—we turned our attention to the impact of program cache size on validator performance. Specifically, we investigated throughput (measured in transactions per second and the validator’s ability to keep up with Solana mainnet-beta) and memory usage (including major and minor page faults, as well as virtual and resident set sizes).

In this section, we shall attempt to answer the following questions:
\begin{enumerate}
    \item Does increasing the program cache increase the demand of the validator on hardware?
    \item If so, does the demand result in an OOM condition?
    \item Can we get faster execution by increasing the program cache?
\end{enumerate}

\subsection{Methodology}
We have evaluated the following cache sizes: 512\footnotemark,~1,024, and~2,048 entries on a machine running Agave of the version v2.1.14.
As these experiments were conducted later, we have chosen the most up-to-date version of the Agave validator for that date.
We chose Latitude.sh’s \code{rs4.metal.xlarge} bare metal servers for this experiment, the hardware specifications for which can be found in Figure~\ref{tab:latitude-hardware}.  
We have chosen a bare metal hosting to eliminate as many variables as possible. 
\footnotetext{%
The default size after v2.1.0, with the previous default being 256.
}
\begin{wrapfigure}{r}{4.2cm}
    \footnotesize
    \begin{tabular}{l p{3cm}}
    \toprule
    \multicolumn{2}{l}{\textbf{CPU}} \\
    & AMD 9554P, \\&64 Cores, \\&3.1 GHz\\\midrule
    \multicolumn{2}{l}{\textbf{RAM}} \\
    & \num{1536} GB \\\midrule
    \multicolumn{2}{l}{\textbf{Storage} } \\
    & 2 \(\times\) 480~GB~NVME  \\
    & 4 \(\times\) 8TB~NVME\footnotemark \\\midrule
    \multicolumn{2}{l}{\textbf{Network} }\\
    & 2 \(\times\) 100~Gbps \\
    \bottomrule
    \end{tabular}
    \caption{Program cache  test machine configuration.}
    \label{tab:latitude-hardware}
\end{wrapfigure}
\footnotetext{We only used two of the~8~TB NVMEs, one for accounts and one for the ledger}

\subsubsection{Metrics Measured}

We measure the following quantities:

\textbf{Virtual Memory Size (VSZ)} – which is the total size of the validator process in memory, including RAM (RSS) and swap.

\textbf{Resident Set Size (RSS)} - which is the portion of a process’s memory that is physically loaded into RAM.

VSZ and RSS is tracked to investigate how much memory is being used by the validator to run. 
Swap utilization can be calculated by subtracting RSS from VSZ. 
Larger utilization of swap indicates less efficient usage of memory.

\textbf{Major Page Faults per second (majflt/s)} – This occurs when a process requests a page that is not in RAM, resulting in a disk access.

\textbf{Minor Page Faults per second (minflt/s)} – This occurs when a process needs to map a page that is already in RAM to its own address space. This does not require disk access.

\textbf{Transactions per Second (TPS)} – the processing throughput of the validator.

Page faults are tracked to show how efficiently the validator uses memory. 
Larger cache sizes, in addition to resulting in faster execution, should require fewer requests to storage. 
A small cache that evicts data from itself and then needs to recall it will require disk access and therefore slow execution; a larger cache will not need to evict anything and will always have what is needed once data is loaded in the cache. On the other hand, too large of a cache could waste space in memory.

TPS is tracked like in any other experiment we run. Beyond answering the question of whether the validator is keeping up with the tip of the chain (and how far behind it is falling if it is not), it is a proxy that allows us to monitor the overall health and connectivity of the validator.

\subsection{Memory Analysis}
First and foremost, we observed that all runs kept up with the chain and had healthy performance characteristics. This was observed by monitoring the validator during its run, and then verifying the TPS data after the runs. The average TPS numbers are consistent with the average Solana throughput. The validaters completed their bootstrapping and managed to catch-up to the chain within~2 hours.

\begin{figure}[tbp]
    \centering
    \begin{subfigure}{\textwidth}
        \centering
        \includegraphics[width=.75\linewidth]{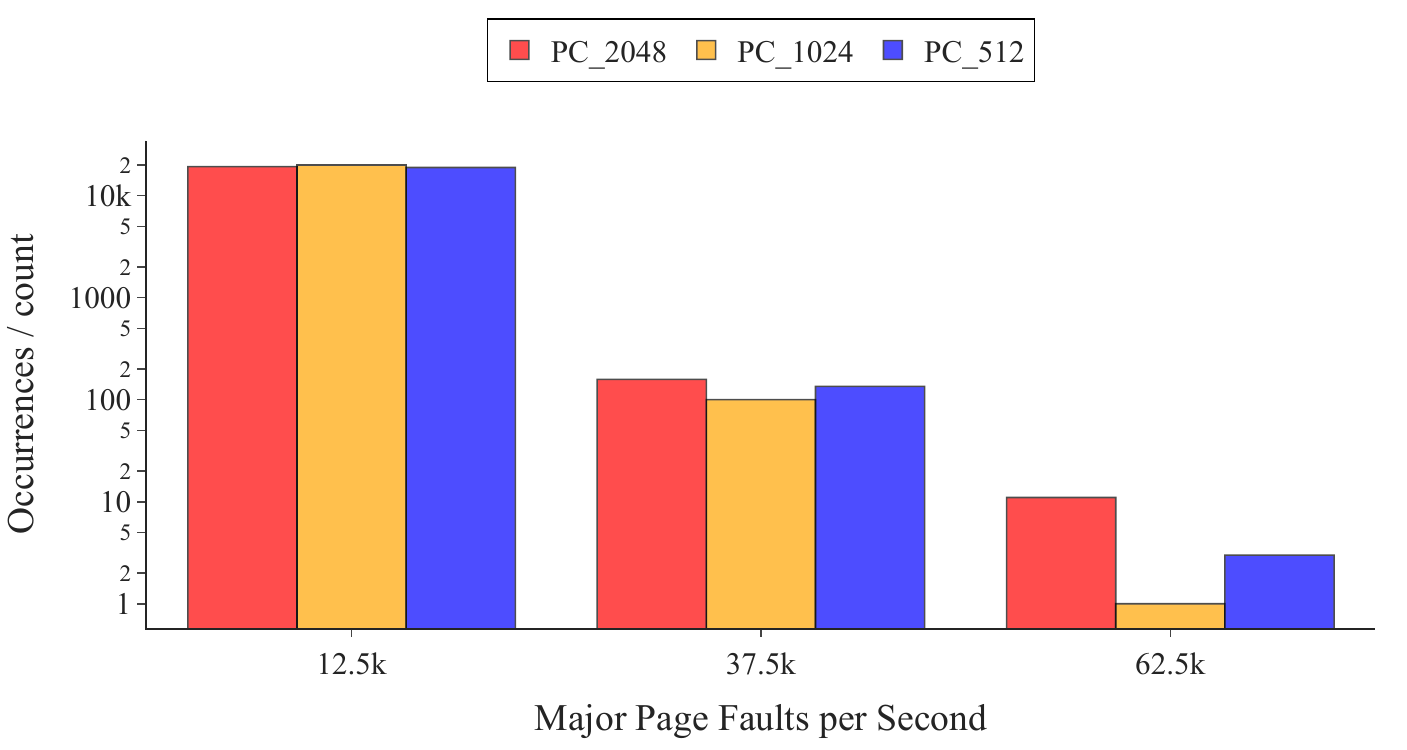}
        \caption{Major}
        \label{fig:majhist}
    \end{subfigure}
    \begin{subfigure}{\textwidth}
        \centering
        \includegraphics[width=.75\linewidth]{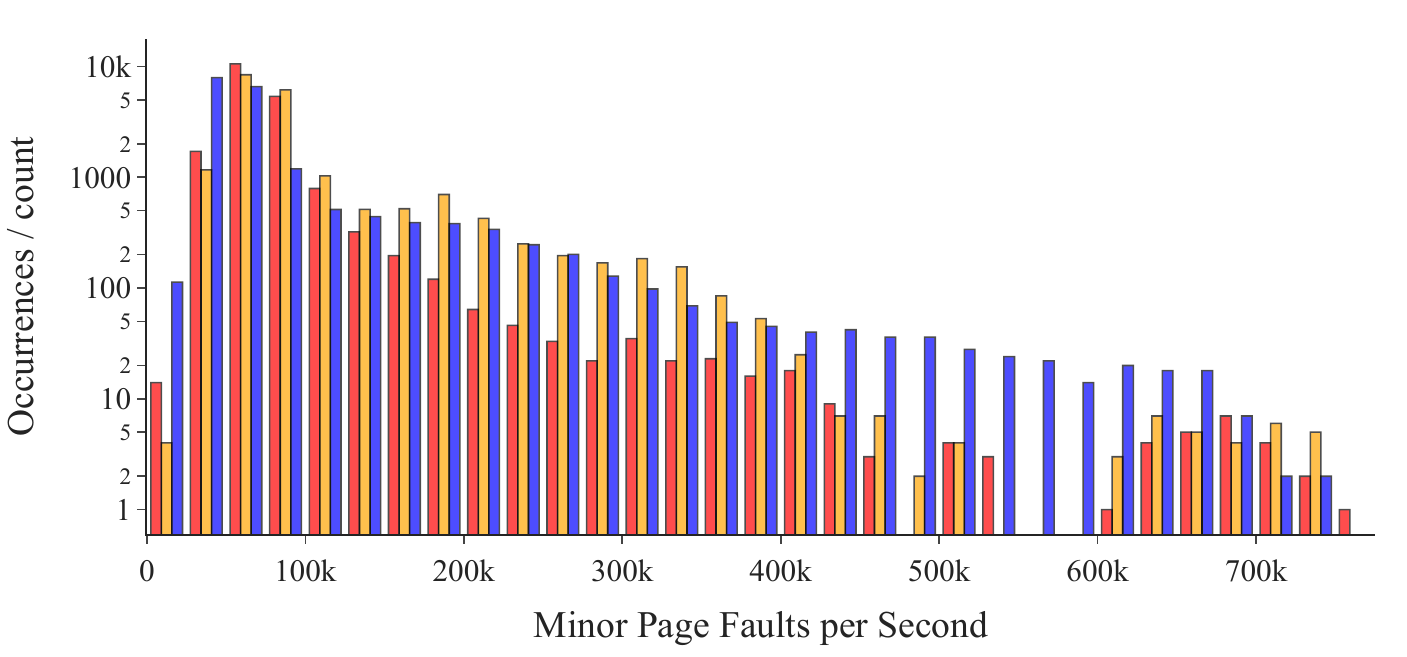}
        \caption{Minor}
        \label{fig:minhist}
    \end{subfigure}
    \caption{A comparison of histograms of different occurrences by type of fault. The major faults are infrequent. The minor faults have a significant number of occurrences, but correlated positively with a larger number of cache entries in the program cache.}
    \label{fig:fault-hist}
\end{figure}

No significant changes in memory were observed during the three runs. Although the patterns of faults look somewhat different, a histogram of minor and major page faults in Figure~\ref{fig:fault-hist} and the average faults show no correlation with the program cache size. Given the RAM size of approximately~1.5~TB, we saw similar memory usage in these runs as the previous experiments with~1 and~1.5~TB of RAM.

Finally, averages across memory metrics show no noticeable correlation between memory usage and program cache size, indicating that the increase in the program cache had minimal impact on the validator performance and memory. The~2,048-entry program cache validator actually used less RAM on average than the~512-entry program cache validator, while the~1,024-entry program cache validator used the most.
We attribute this to circadian cycles of demand, not to increased memory pressure.
As such having a larger JIT cache offers minimal downsides, unless the machine is significantly memory-starved.

\begin{figure}[tb]
\centering\footnotesize
\begin{tabular}{lrrr}
\toprule
\textbf{Cache entries} & {2,048} & {1,024} & {512} \\
\midrule
\textbf{RSS} & & & \\ \midrule
Min  & \num{0.016}  & \num{0.016}  & \num{0.016} \\

Max  & \num{607.933}   & \num{619.631}   & \num{634.235} \\

Mean & \num{476.262}   & \num{501.673}   & \num{489.175} \\
\midrule
\textbf{VSZ} & & & \\ \midrule
Min  & \num{0.198}  & \num{0.100}  & \num{0.111} \\
Max  & \num{767.905}   & \num{804.717}   & \num{798.045} \\
\bottomrule
\end{tabular}
\caption{Memory usage statistics by Program Cache (PC) configuration. The number is given in the number of entries. }
\label{tab:pc-memory}
\end{figure}

\begin{figure}[tbp]
    \centering
    \includegraphics[width=.75\linewidth]{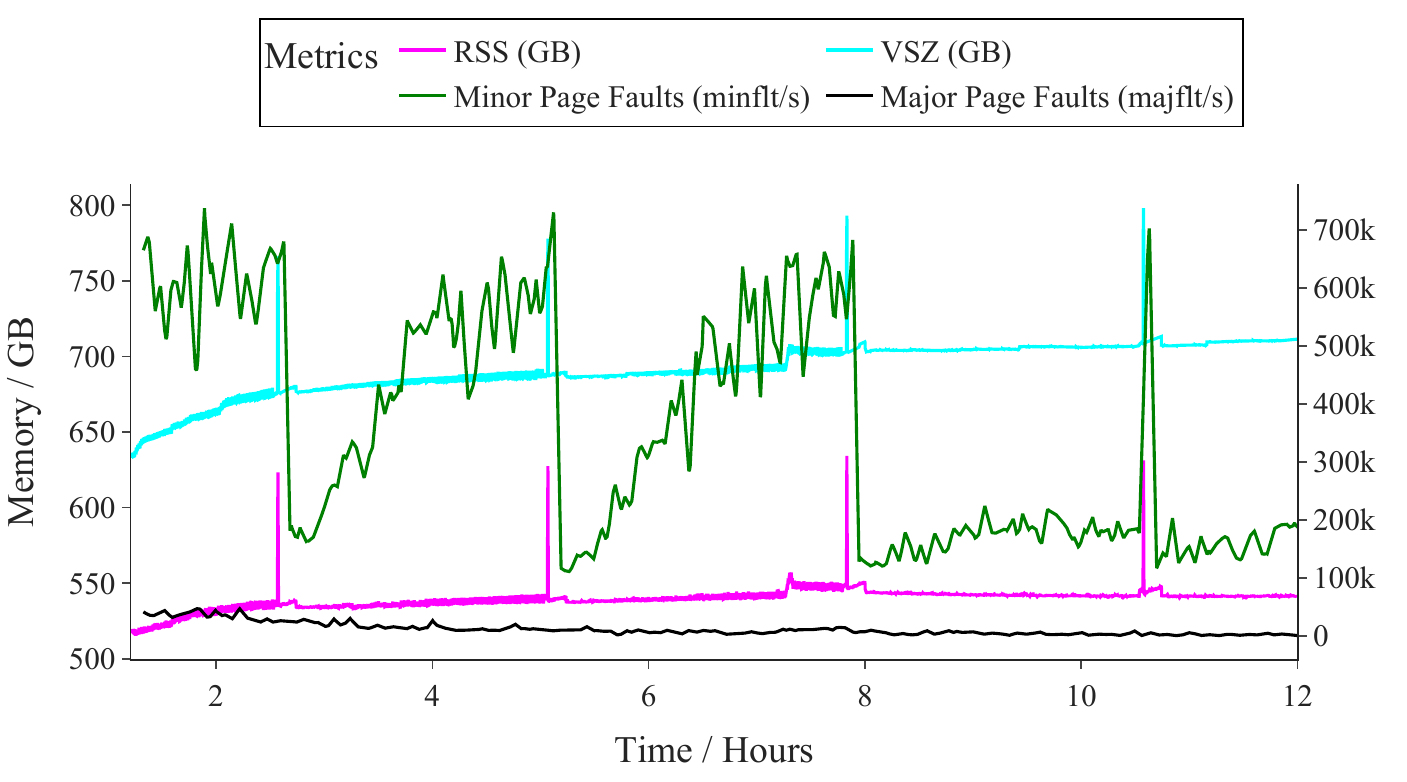}
    \caption{Correlation of RSS, VSZ, Major and Minor page faults for the 512 entries PC}
    \label{fig:pc_mem_comp}
\end{figure}

\begin{figure}[tbp]
    \centering
    \begin{subfigure}{.75\textwidth}
        \centering
        \includegraphics[width=\linewidth]{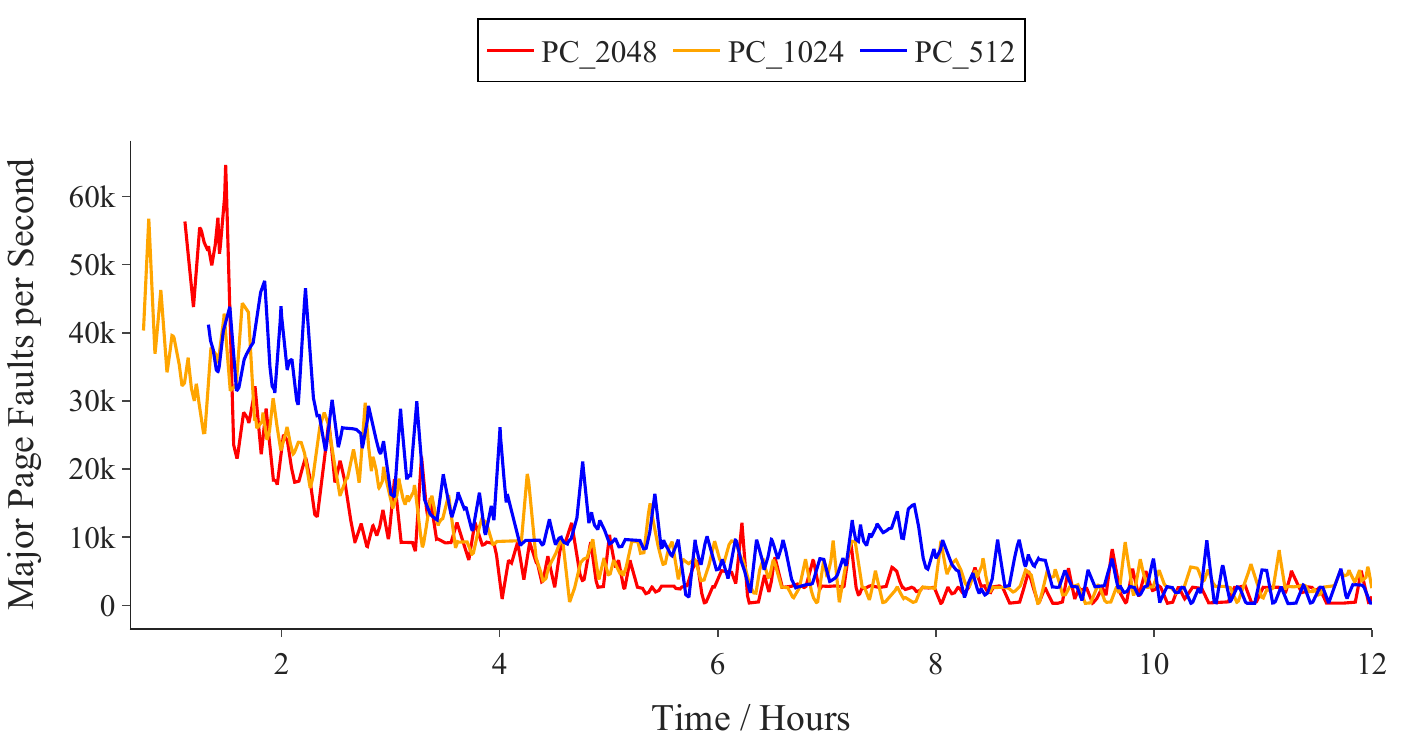}
        \caption{Major}
        \label{fig:majfault}
    \end{subfigure}
    \begin{subfigure}{.75\textwidth}
        \centering    
        \includegraphics[width=\linewidth]{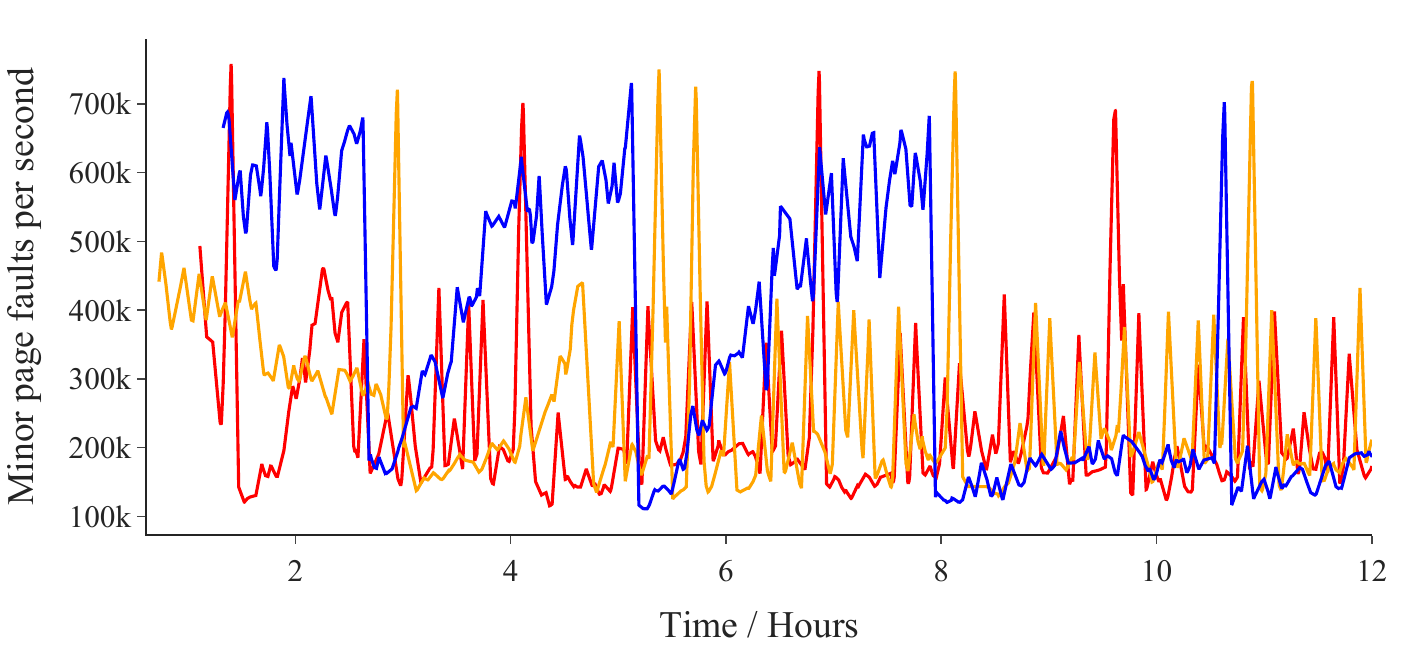}
        \caption{Minor}
        \label{fig:minfault}
    \end{subfigure}
    \caption{%
    A timeline of major (a) and minor (b) page faults. 
    Note that while the major page faults level off at approximately~12 hour mark, the number of minor page faults remains significant. 
    Data was smoothed with a sliding window of~100 entries.
    }%
    \label{fig:fault_time}
\end{figure}

\begin{figure}[tbp]
    \centering
    \includegraphics[width=.75\linewidth]{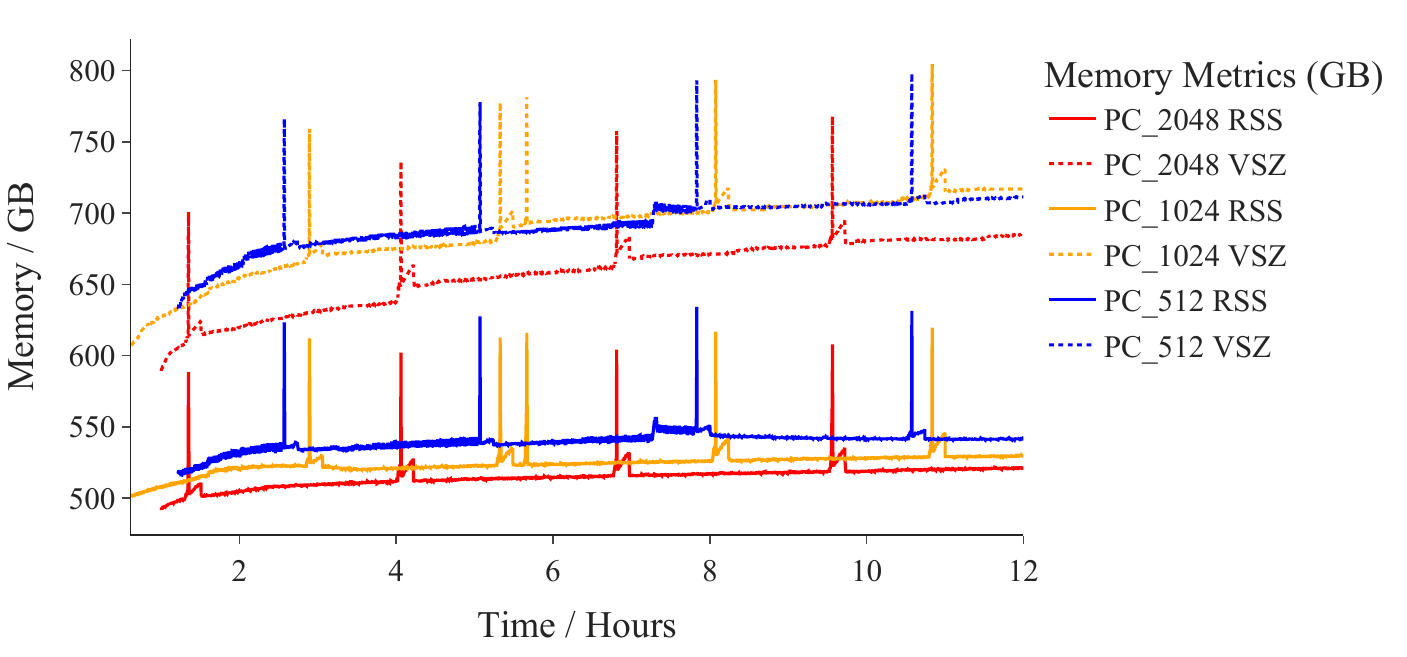}
    \caption{A comparison of the memory usage between different program cache sizes. The spikes do not correspond to one another, because the data was not taken concurrently.}
    \label{fig:rss-vsz}
\end{figure}


\subsection{Cache and TPS Analysis}
When we analyze execution times and cache performance, however, we observe a significant performance benefit. 
During a~12-hour run, the aggregate latency contributed by the program cache amounted to approximately one hour. 
This was brought down to~28 minutes and~23 minutes by the~1,024- and~2,048-entry program caches, respectively: we were able to reduce the latency of the program cache $\tfrac{2}{3}$ in the case of the~2,048-size cache, and $\tfrac{1}{2}$ in the case of the~1,024-size cache.

This is reflected in the average time spent in the program cache per execution: down from~40~ms on average spent in the program cache per execution to~15 and~13~ms. This is explained neatly by the number of missed and evictions across the cache sizes. While the standard cache consisting of~512-entries had an average of~0.84 evictions per execution and~0.85 misses (with a maximum of~134 misses, requiring~134 subsequent loads and no more than~12 evictions to accommodate), the~2,048-entry cache was able to handle the entirety of execution over the~12 hour period without needing a single eviction, while the~1,024-entry cache had~90\% fewer misses and evictions of the~512-entry cache. This could even imply that with the current diversity of on-chain programs, we can have a cache that can hold all commonly executed programs, while not increasing the load on validator hardware.

Finally, we included the prune time to ensure that, although the cache may lead to speedups, it would also not require more time to maintain. 
The time spent pruning the program cache (similar to pruning dead forks during execution) did not increase.
By contrast, there was a slight downtrend in pruning time as the size of the cache increased, indicating that the larger cache is not only faster, but just as light and perhaps even lighter when with respect to keeping consistency with the chain.

\begin{figure}
    \centering
    \includegraphics[width=.75\linewidth]{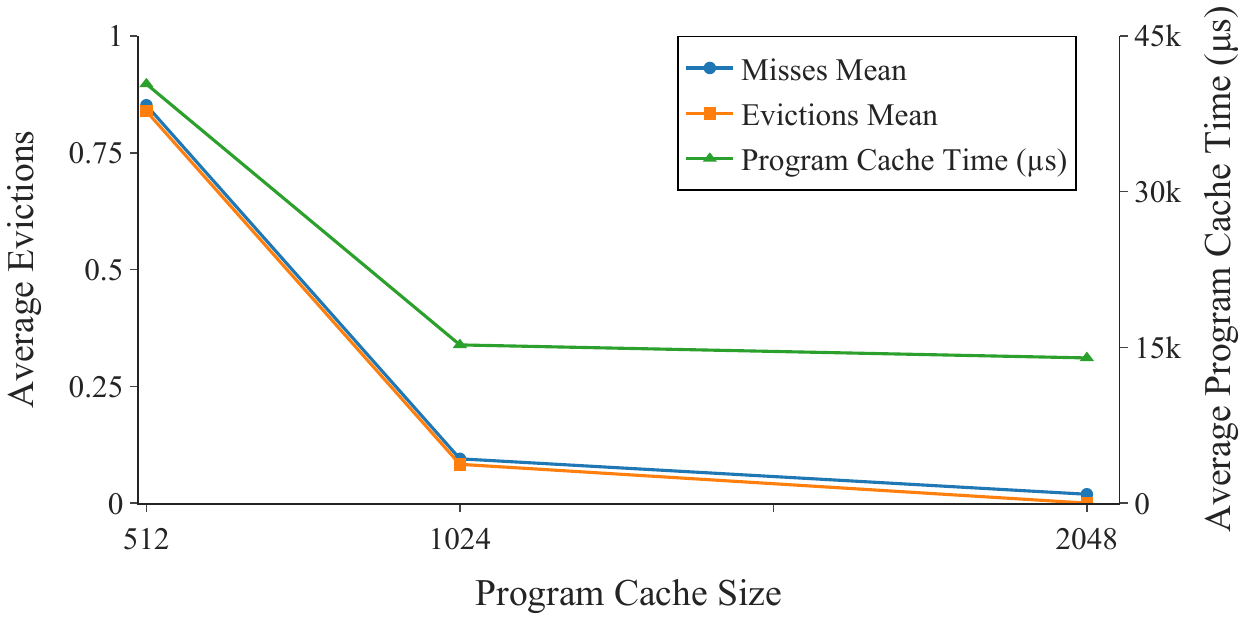}
    \caption{Average program cache misses, evictions and time spent in program cache per execution batch as number of program cache entries increase.}
    \label{fig:cache_comparison}
\end{figure}

\begin{figure}[tb]
\centering\footnotesize
\begin{tabular}{l r r r}
\toprule
\textbf{Metric} & \textbf{2,048} & \textbf{1,024} & \textbf{512} \\
\midrule
Misses (Sum)            & \num{1925}     & \num{10538}    & \num{86841} \\
Misses (Mean)           & \num{0.02}      & \num{0.09}      & \num{0.85} \\
\midrule
Evictions (Sum)         & \num{0}         & \num{9233}     & \num{85530} \\
Evictions (Mean)        & \num{0.00}      & \num{0.08}      & \num{0.84} \\
\midrule
PC Time ($\mu$ s) (Sum)      & \num{1417951905} & \num{1692786514} & \num{4114784006} \\
PC Time ($\mu$ s) (Mean)     & \num{13984.30} & \num{15239.21} & \num{40363.97} \\
\bottomrule
\end{tabular}
\caption{Program Cache statistics across the configurations. The sum is the aggregate over all data points.}
\label{tab:program-cache}
\end{figure}

\begin{figure}
    \centering
    \begin{subfigure}{\textwidth}
        \centering
        \includegraphics[width=0.6\linewidth]{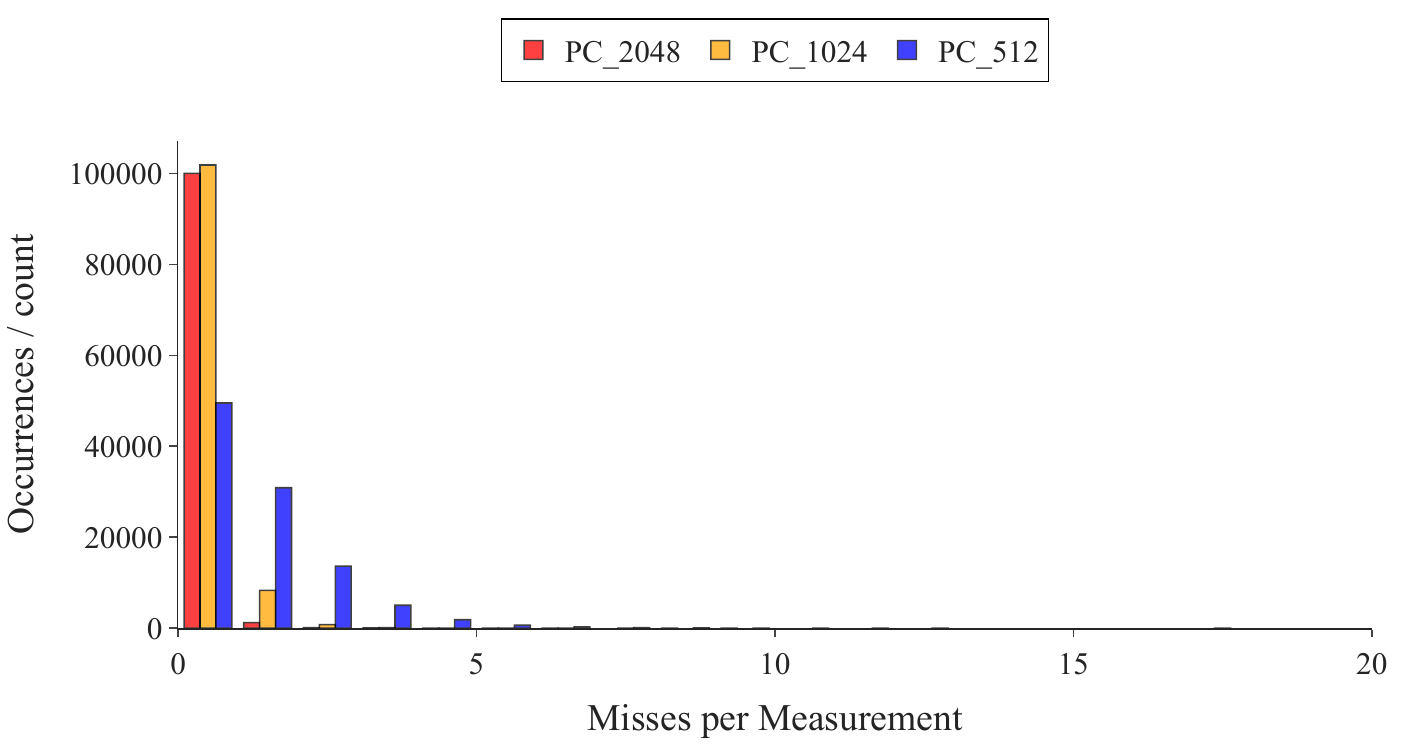}
        \caption{Histogram of program cache misses per execution batch.}
        \label{fig:misses-histogram}
    \end{subfigure}
    \begin{subfigure}{\textwidth}
        \centering
        \includegraphics[width=0.6\linewidth]{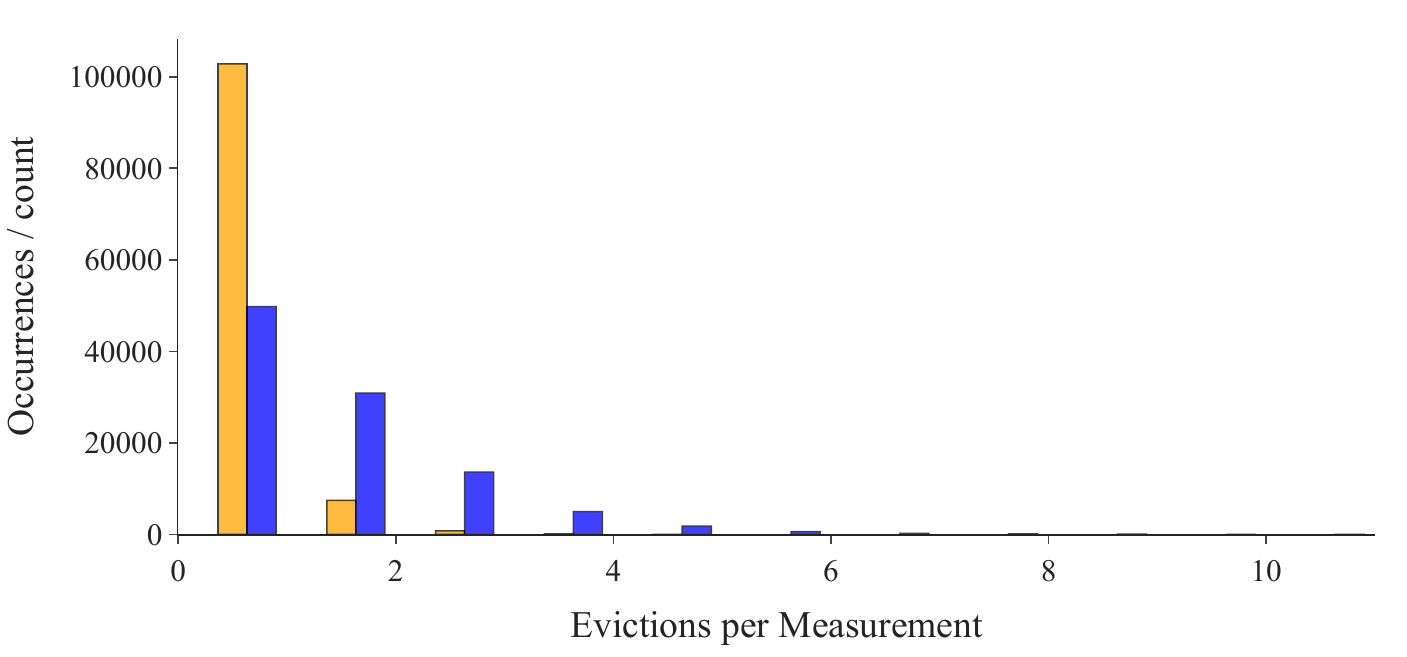}
        \caption{Histogram of evictions of program cache entries per execution batch.}
        \label{fig:evictions-histogram}
    \end{subfigure}
    \begin{subfigure}{\textwidth}
        \centering
        \includegraphics[width=0.6\linewidth]{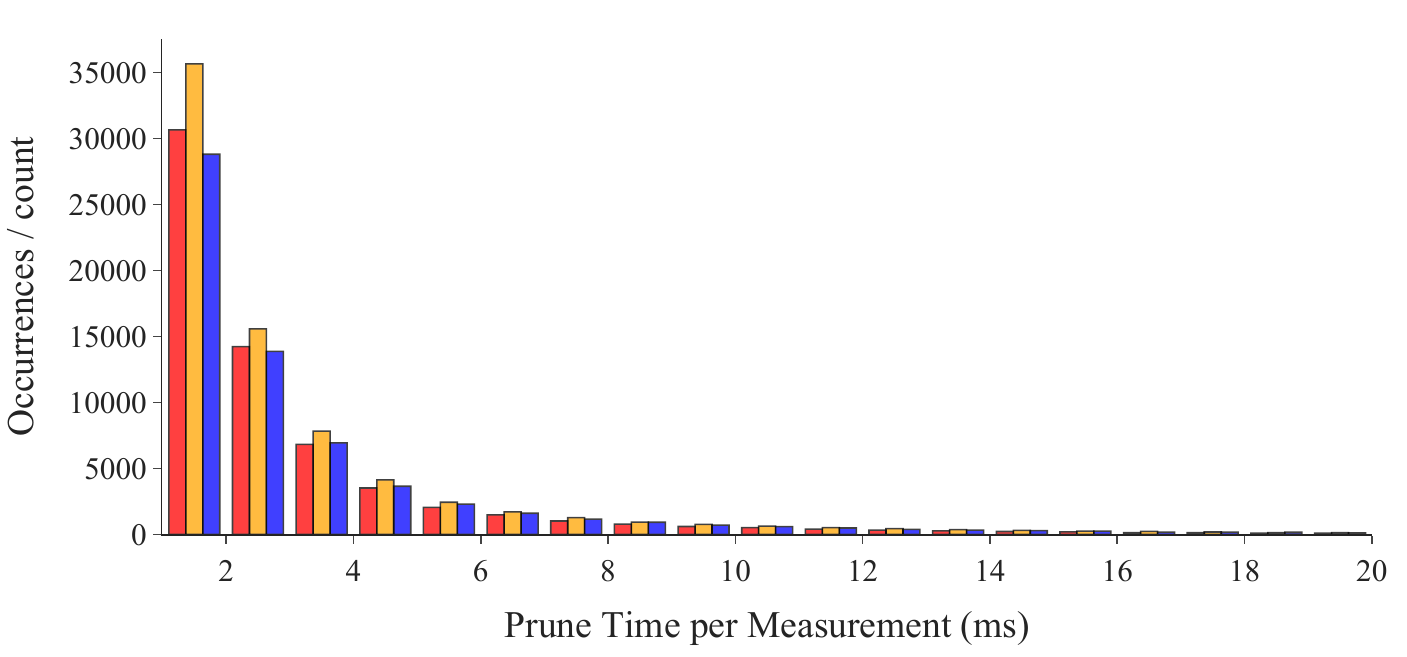}
        \caption{Histogram of time needed to prune the program cache per execution batch.}
        \label{fig:prune-histogram}
    \end{subfigure}
    \begin{subfigure}{\textwidth}
        \centering
        \includegraphics[width=0.6\linewidth]{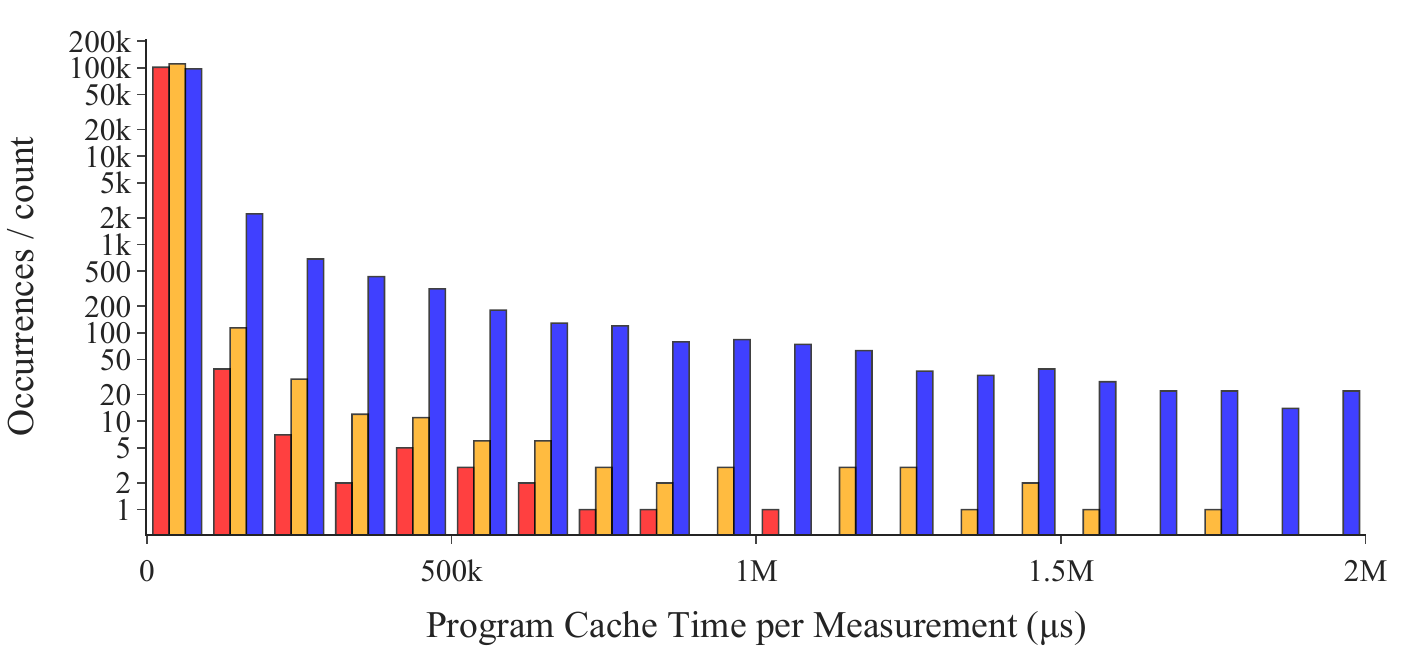}
        \caption{Histogram of time spent in the program cache per execution batch.}
        \label{fig:pc_time_histogram}
    \end{subfigure}
    \caption{A comparison of program cache performance by evictions, misses, prune time, and time spent in the program cache per execution batch. Graphs are truncated at higher x values for legibility. Measurements show similar prune time across all cache sizes, however much more efficient program cache performance as cache size increases.}
    \label{fig:pc_analysis}
\end{figure}

\subsection{Discussion}
Our validator was able to keep up with the chain and run uninhibited by the varying program cache sizes, in all but one configuration (see footnote for Figure~\ref{tab:pc-memory}). 
There was no significant increase in memory demand from the larger cache sizes, but the validator was able to spend fewer cycles and less time in the cache during execution. We were able not only to get to a cache that holds all the programs needed in execution over a~12 hour period with~0 evictions, we were able to bring evictions down from~0.85 per execution on average to~0.09 and then further to~0.02 by simply doubling and then quadrupling the baseline cache size.

\section{Related Work}%
\label{sec:related-work}


We begin this section by pointing out that there is comparatively little work done on the subject of RAM usage for the Solana validator specifically. Most of the attention is dedicated to Ethereum and is thus of limited relevance.  We present a few measurements that we found to be useful. 

\point{Ethereum}
We begin with Marzouqi~\etal~\cite{Marzouqi}, who have studied the effects of memory and computational constraints on the performance of GEth, an implementation of the Ethereum blockchain in Go. 
Their work demonstrated that Ethereum suffers greatly in a constrained environment.
Not only was the performance far inferior on the Raspberry Pi, but the PC-to-PC benchmark showed much better power consumption, which they attributed to consensus.

We believe that a more systematic approach, as in~\cite{ehtereum-2-0-clients} would benefit the SVM ecosystem.  
Different implementations of the EVM showed a significant variety of performance characteristics, which would be beneficial to study in the SVM space. 
However, at the moment, there are not enough completed implementations of the SVM, with significant differences.
As of writing, for example, the Firedancer client statically links elements of the Agave validator. 
Similarly, Jito only offers a thin layer of patches on top of Agave, severely limiting the variety of blockchain clients.
A follow-up study examining the differences between these clients would be beneficial once the said clients are complete.

Ethereum has more types of data that must be persisted to disk than Solana.  
This manifests as the persistent storage often being cited as the prime suspect of poor performance for an Ethereum node. 
Specifically, Yang~\etal~\cite{yangSolsDBSolveEthereums2024} identified the bottleneck of Ethereum as due to write amplification. 
They proposed a single-level ordered log structure database (SolsDB), which stores state data in blocks using a globally ordered file structure, eliminating write amplification caused by compaction. 
Its Parser module interprets query keys to enable direct file access without level-by-level searching, solving the read amplification problem. 
Compared to LevelDB, it achieves up to~$4.7\times$ faster read speeds, reduces read tail latency by~68.7\% to~83.3\%, and decreases write amplification by~49.1\% to~76.1\%. 

Other studies, \textit{e.g.} \cite{jezekEthereumDataStructures2021,cortes-goicoecheaResourceAnalysisEthereum2020,masoudMeasurementStudyEthereum2024}, took a more detailed look at the data structures used in the Ethereum code base and compare them with industry standards.  
This is an approach that we would like to adopt in a follow-up investigation.  Toyoda~\etal~\cite{toyodaFunctionLevelBottleneckAnalysis2020} have taken the investigation further by analyzing the performance traces using Go's standard tools, an approach that we could adopt ourselves. 
A deeper look into consensus in terms of finality and fork choice is also a worthwhile method to explore as a follow-up, as was done in, for example, Murr~\etal~\cite{murrSingleSlotFinality2024}.

\point{Solana}
There are a limited number of case studies conducted on Solana's ability to scale. Pierro~\etal~\cite{pierroCanSolanaBe2022}, took a principled approach and confirmed data by monitoring Solana itself.  
They have confirmed that the priority fees on Solana were much lower than other blockchains and, indeed, that Solana was operating much below its theoretical limits at~3,000~TPS over that time frame. 

Follow up work by Lebedev~\etal~\cite{lebedevRelevanceBlockchainEvaluations2023} focused on identifying the relative importance of using bare metal hosting compared to container-based cloud infrastructure.  
They show that LAN switches had minimal impact on the latency, but also identified that the latency reported by the blockchains was not a reliable proxy for indicating the tail latency of transaction processing. 
There have been a number of similar studies, \textit{e.g.} Dinh~\etal~\cite{dinh} showcase Blockbench, a benchmarking framework designed with a focus on permissioned blockchains. 
They use a commodity cluster of~48 machines interconnected with a gigabit switch. 
In the experiments, the number of blockchain nodes ranged from~1 to~32. 
These approaches can be extended, to our end goal, but fail to capture the behavior of the system under natural load. 
Finally, a study conducted by Diamandis~\etal~\cite{diamandisDesigningMultidimensionalBlockchain2023} investigates the effects of setting priority fees correctly for the system.

\section{Conclusions}%
\label{sec:conclusion}

Solana is an efficient protocol.
Careful design of caches and pipelines spanning the TPU, as well as the design of the bank and the SVM result in Solana having the best performance both in terms of transactions per second and efficient utilization of the hardware.
The trend of improving performance and validator stability is continued with Agave.
In this paper we analyzed the impact of hardware choices on the performance of the Agave validator, confirming that the software is not the limiting factor.

The interplay between execution, cache, and system resources is an active area of research, with the potential for substantial performance gains through more intelligent memory management and algorithmic refinements. As Solana continues to evolve, ongoing experimentation and analysis will be crucial in unlocking new levels of scalability and efficiency in its execution model.
We thank Anza for providing some valuable input regarding this work. 

Our analysis of the memory tunable parameters indicates that Agave is stable on machines with 512~GiB of RAM allocated to the validator process.
Configurations in excess of a terabyte of RAM, offer diminishing returns.
By interpolating our data, we believe that it is not worthwhile to pursue configurations in excess of approximately 700~GiB. 
We also found that it is at present not viable to choose a configuration below 256~GiB, with 128~GiB presenting the most likely failure mode: inability to keep up with the network, OOM crashes \textit{etc.}

Not all of our findings necessarily require better hardware.
We found that increasing the size of the program cache avoids recompilation of programs accessed within a 12-hour window.
Increasing the size of the program cache to 2048 entries resulted in an overall 90\% reduction of the latency contributed by the program cache operations. 

Our work further highlights the relative vacuum and scarcity of investigations into Solana's tunable parameters.
As a follow up work, we propose the following investigations:
\begin{enumerate}
    \item Further evaluation of Agave's caches: Performing an analogous investigation of the accounts cache.  
    \item A rolling analysis of memory allocations: Performing one such study imposes a significant time delay between data collection and publication of findings.  In our case specifically, the issues that we have identified had been reported and fixed during said time.  While our data is largely still applicable, we believe that faster feedback cycles and reporting would be beneficial to the whole ecosystem.
    \item Eviction strategy improvements: Profiling the probabilistic 2’s random eviction method against real-time program usage statistics.  
    It may be worth investigating the trade-offs of maintaining a partially accurate ranking of programs by usage frequency, and implementing a complete LFU eviction.
    While this may not be suitable for node operators due to the potential of remote code execution bugs exposed via the cache, we believe that comparing the performance may be valueable.
\end{enumerate}
%

\bibliographystyle{plain}


\end{document}